\RequirePackage[l2tabu, orthodox]{nag}
\newcommand*{\ATLASLATEXPATH}{}
\documentclass[USenglish,texlive=2011,PAPER,cernpreprint]{\ATLASLATEXPATH atlasdoc}
\pdfoutput=1
\PreprintIdNumber{CERN-PH-EP-2016-001}

\usepackage[full]{\ATLASLATEXPATH atlaspackage}
\usepackage{comment}
\usepackage{\ATLASLATEXPATH atlasbiblatex}
\usepackage{overpic}
\usepackage{comment}

\usepackage{\ATLASLATEXPATH atlascontribute}

\usepackage{\ATLASLATEXPATH atlasphysics}
\graphicspath{{}}

\usepackage{multirow}

\hypersetup{pdftitle={ATLAS draft},pdfauthor={The ATLAS Collaboration}}
\graphicspath{{logos/}}
\usepackage{booktabs,siunitx}
\usepackage{tikz}
\usetikzlibrary{automata,positioning}

\AtlasTitle{Measurement of the charged-particle multiplicity inside jets from $\sqrt{s}=8$ \TeV~$pp$ collisions with the ATLAS detector}

\AtlasJournal{EPJC}

\AtlasAbstract{%
The number of charged particles inside jets is a widely used discriminant for identifying the quark or gluon nature of the initiating parton and is sensitive to both the perturbative and non-perturbative components of fragmentation.  This paper presents a measurement of the average number of charged particles with $p_\text{T}>500$ \MeV~inside high-momentum jets in dijet events using 20.3 fb$^{-1}$ of data recorded with the ATLAS detector in $pp$ collisions at $\sqrt{s}=8$~\TeV~collisions at the LHC.  The jets considered have transverse momenta from 50~\GeV~up to and beyond 1.5 \TeV.  The reconstructed charged-particle track multiplicity distribution is unfolded to remove distortions from detector effects and the resulting charged-particle multiplicity is compared to several models.  Furthermore, quark and gluon jet fractions are used to extract the average charged-particle multiplicity for quark and gluon jets separately. 
}

\begin{document}
\maketitle

%	\begin{center}
%	\plot{X axis label}{Y axis label}{Some text. Don't worry about wrapping...tex should take care of that for you.}
%	\end{center}	

\section{Introduction}
\label{sec:intro}

Quarks and gluons produced in high-energy particle collisions hadronize before they can be observed directly.  However, the properties of the resulting collimated sprays of hadrons, known as jets, depend on the type of parton which initiated them.  One jet observable sensitive to the quark or gluon nature is the number of charged particles inside the jet.  Due to their larger colour-charge under the strong force, gluon-initiated jets contain on average more particles than quark-initiated jets.  The average (charged) particle multiplicity inside jets increases with jet energy, but increases faster for gluon-initiated jets than for quark-initiated jets~\cite{pinkbook}.   These properties were used recently at the Large Hadron Collider (LHC) to differentiate between jets originating from a quark or a gluon~\cite{Aad:2014gea,Aad:2015owa,Aad:2014bia,Khachatryan:2015bnx,Khachatryan:2014dea}. These studies have found significant differences in the charged-particle multiplicity between the available simulations and data.  Improved modelling based on measurements of the number of charged particles inside jets is thus crucial for future studies.

This paper presents a measurement of the average charged-particle multiplicity inside jets as a function of the jet transverse momentum in dijet events in $pp$ collisions at $\sqrt{s}=8$ \TeV~with the ATLAS detector.  The measurement of the charged-particle multiplicity inside jets has a long history from the SPS~\cite{Arnison:167874,BAGNAIA1984291,ua22},  PETRA~\cite{Bartel:1982ub,zphys}, PEP~\cite{Derrick:163229,PhysRevLett.55.1954,PhysRevLett.72.3145,Abe:1996zi}, TRISTAN~\cite{PhysRevLett.63.1772}, CESR~\cite{PhysRevD.56.17}, LEP~\cite{Alexander:1991ce,Acton:1993jm,xyz,Buskulic1995389,Alexander1996659,Buskulic1996353,delphintrack,refId0,Abreu1999383,refId1,Akrawy1990617,opallep2}, and the Tevatron~\cite{Affolder:2001jx}.  At the LHC, both ATLAS~\cite{Aad:2011gn,Aad:2011sc} and CMS~\cite{Chatrchyan:2012mec} have measured the charged-particle multiplicity inside jets at $\sqrt{s}=7$ \TeV.  One ATLAS result~\cite{Aad:2011gn} used jets that are reconstructed using tracks and have transverse momentum less than 40 \GeV.   A second ATLAS analysis~\cite{Aad:2011sc} has measured charged particles inside jets with transverse momenta spanning the range from 50 to 500 \GeV with approximately constant 3--4\% uncertainties.  The CMS measurement~\cite{Chatrchyan:2012mec} spans jet transverse momenta between 50 and 800 \GeV with 5--10\% uncertainties in the bins of highest transverse momentum.  The analysis presented here uses the full $\sqrt{s}=8$ \TeV ATLAS dataset, which allows for a significant improvement in the precision at high transverse momentum up to and beyond 1.5 \TeV.  

This paper is organized as follows.  After a description of the ATLAS detector and object and event selection in Sect.~\ref{sec:atlas}, simulated samples are described in Sect.~\ref{sec:samples}.  In order for the measured charged-particle multiplicity to be compared with particle-level models, the data are unfolded to remove distortions from detector effects, as described in Sect.~\ref{sec:particlelevel}.  Systematic uncertainties in the measured charged-particle multiplicity are discussed in Sect.~\ref{sec:uncerts} and the results are presented in Sect.~\ref{sec:results}.  

\section{Object and event selection}
\label{sec:atlas}

ATLAS is a general-purpose detector designed to measure the properties of particles produced in high-energy $pp$ collisions with nearly a full $4\pi$ coverage in solid angle.  Charged-particle momenta are measured by a series of tracking detectors covering a range\footnote{ATLAS uses a right-handed coordinate system with its origin at the nominal interaction point (IP) in the centre of the detector and the $z$-axis along the beam pipe. The $x$-axis points from the IP to the centre of the LHC ring, and the $y$-axis points upward. Cylindrical coordinates $(r,\phi)$ are used in the transverse plane, $\phi$ being the azimuthal angle around the beam pipe. The pseudorapidity is defined in terms of the polar angle $\theta$ as $\eta=-\ln\tan(\theta/2)$.  The variable $\Delta R=\sqrt{(\Delta\phi)^2+(\Delta\eta)^2}$ is a measure of how close two objects are in the $(\eta,\phi)$ plane.} of $|\eta|<2.5$ and immersed in a 2 T axial magnetic field, providing measurements of the transverse momentum, $p_\text{T}$, with a resolution $\sigma_{p_\text{T}}/p_\text{T} \sim 0.05\%\times p_\text{T}/\text{ \GeV} \oplus 1\%$. Electromagnetic and hadronic calorimeters surround the tracking detector, with forward calorimeters allowing electromagnetic and hadronic energy measurements up to $|\eta|=4.5$.  A detailed description of the ATLAS detector can be found in Ref.~\cite{Aad:2008zzm}.

This measurement uses the dataset of $pp$ collisions recorded by the ATLAS detector in 2012, corresponding to an integrated luminosity of 20.3 fb${}^{-1}$ at a center-of-mass energy of $\sqrt{s}=8$ \TeV.  The data acquisition and object/event selection are described in detail in Ref.~\cite{Aad:2015cua} and highlighted here for completeness.  Jets are clustered using the anti-$k_t$ jet algorithm~\cite{Cacciari:2008gp} with radius parameter $R=0.4$ implemented in FastJet~\cite{Cacciari:2011ma} using as inputs topological calorimeter-cell clusters~\cite{TopoClusters}, calibrated using the local cluster weighting (LCW) algorithm \cite{EndcapTBelectronPion2002,Barillari:2009zza}.  An overall  jet energy calibration accounts for residual detector effects as well as contributions from multiple proton--proton collisions in the same bunch crossing (pileup)~\cite{areasATLAS} in order to make the reconstructed jet energy correspond to an unbiased measurement of the particle-level jet energy.  Jets are required to be central $(|\eta| < 2.1)$ so that their charged particles are within the $|\eta|<2.5$ coverage of the tracking detector.    Events are further required to have at least two jets with $p_\text{T}>50$ \GeV~and only the leading two jets are considered for the charged-particle multiplicity measurement.  To select dijet topologies where the jets are balanced in $p_\text{T}$, the two leading jets must have $p_\text{T}^\text{lead}/p_\text{T}^\text{sublead} < 1.5$, where $p_\text{T}^\text{lead}$ and $p_\text{T}^\text{sublead} $ are the transverse momenta of the jets with the highest and second-highest $p_\text{T}$, respectively.  The jet with the smaller (larger) absolute pseudorapidity $|\eta|$ is classified as the more central (more forward) jet.  A measurement of the more forward and more central average charged-particle multiplicities can exploit the rapidity dependence of the jet type to extract information about the multiplicity for quark- and gluon-initiated jets as is described in Sect.~\ref{sec:results}.  The more forward jet tends to be correlated with the parton with higher longitudinal momentum fraction $x$, and is less likely to be a gluon-initiated jet.  

\begin{figure}[h!]
\begin{center}
\includegraphics[width=0.8\textwidth]{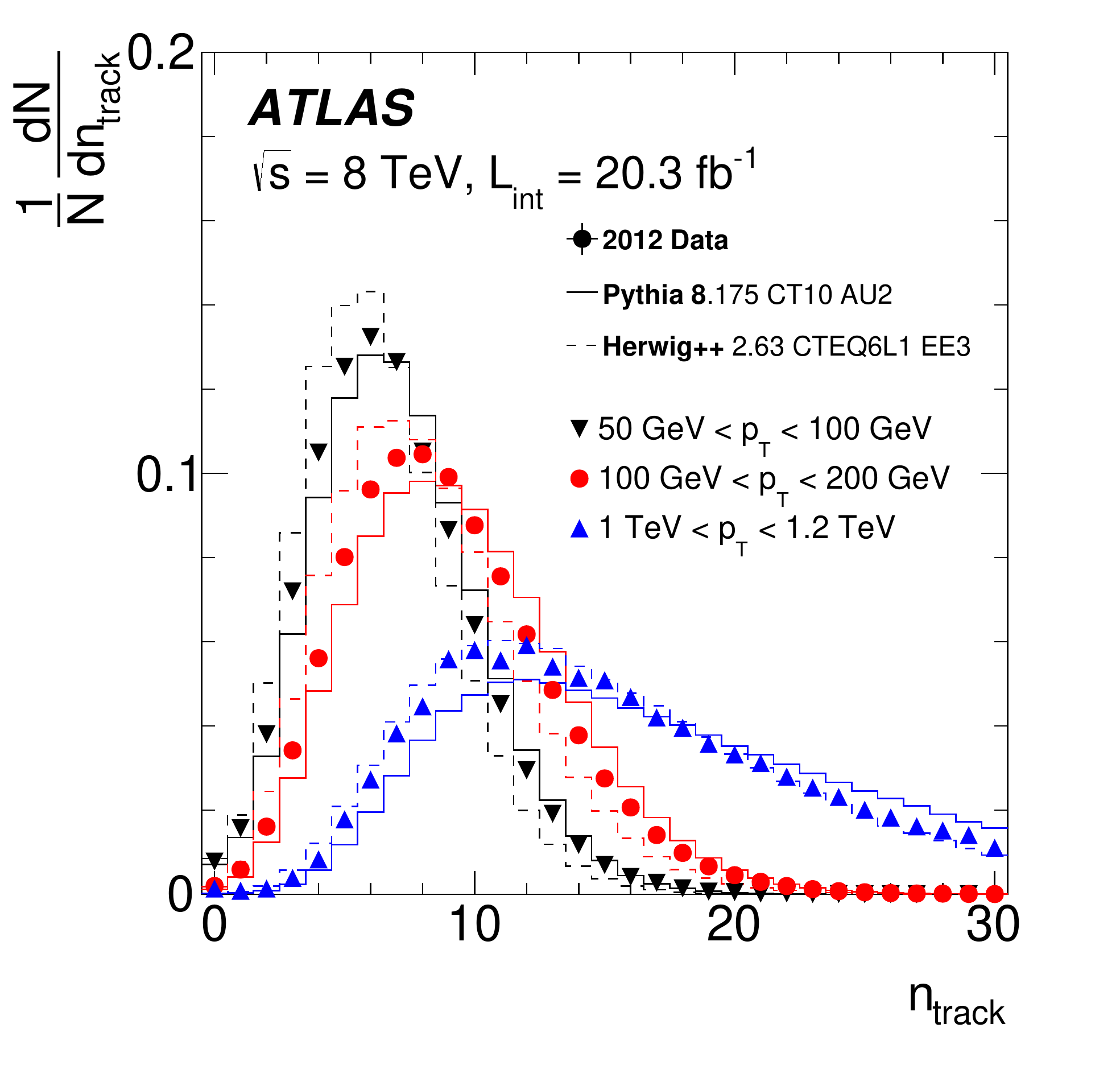}
\end{center}	
\caption{The distribution of the number of reconstructed tracks associated with a jet (not unfolded) in three example jet $p_\text{T}$ ranges: 50 \GeV $<p_\text{T}<$ 100 \GeV, 100 \GeV $<p_\text{T}<$ 200 \GeV, and 1 \TeV $<p_\text{T}<$ 1.2 \TeV~for data and for {\sc Pythia 8} and {\sc Herwig++} predictions.  The simulated samples are described in Sect.~\ref{sec:samples}.  The data points have statistical uncertainties which in all bins are smaller than the marker size.  There is one entry per jet.}
\label{fig:tracks}
\end{figure}

Tracks are required to have $p_\text{T} \geq$ 500 \MeV, $|\eta| < 2.5$,~and a $\chi^2$ per degree of freedom (resulting from the track fit) less than 3.0.    Additional quality criteria are applied to select tracks originating from the collision vertex and reject fake tracks reconstructed from random hits in the detector.  In particular, tracks are matched to the hard-scatter vertex by requiring $|z_0\sin(\theta)|<1.5$ mm and $|d_0|<1$ mm, where $z_0$ and $d_0$ are the track longitudinal and transverse impact parameters, respectively, calculated with respect to the primary vertex.  Tracks must furthermore have at least one hit in the silicon pixel detector and at least six hits in the semiconductor microstrip detector.  The matching of tracks with the calorimeter-based jets is performed via the ghost-association technique~\cite{area}: the jet clustering process is repeated with the addition of `ghost' versions of measured tracks that have the same direction but infinitesimally small $p_\text{T}$, so that they do not change the properties of the calorimeter-based jets. A track is associated with a jet if its ghost version is contained in the jet after reclustering.  The distribution of the number of tracks in three representative jet $p_\text{T}$ ranges is shown in Fig.~\ref{fig:tracks}.  The number of tracks increases with jet $p_\text{T}$ and the data fall mostly between the distributions predicted by {\sc {\sc Pythia}} and {\sc {\sc Herwig++}} Monte Carlo simulations.

\clearpage

\section{Event simulation}
\label{sec:samples}

Monte Carlo (MC) samples are used in order to determine how the detector response affects the charged-particle multiplicity and to make comparisons with the corrected data. The details of the samples used are shown in Table~\ref{tab:mc_samples1}.  The sample generated with {\sc Pythia} 8.175~\cite{Sjostrand:2007gs} using the AU2~\cite{ATL-PHYS-PUB-2012-003} set of tuned parameters (tune) and the {\sc Herwig++} 2.6.3~\cite{Bahr:2008pv} sample with the UE-EE3~\cite{Gieseke:2012ft} tune are further processed with the ATLAS detector simulation~\cite{Aad:2010ah} based on GEANT4~\cite{Agostinelli:2002hh}.  The effects of pileup are modelled by adding to the generated hard-scatter events (before the detector simulation) multiple minimum-bias events generated with {\sc Pythia} 8.160, the A2 tune~\cite{ATL-PHYS-PUB-2012-003}, and the MSTW2008LO~\cite{Martin:2009iq} Parton distribution function (PDF) set.  The distribution of the number of interactions is then weighted to reflect the pileup distribution in the data.

\begin{table}[h!]
\centering
\begin{tabular}{ccccccc}
ME Generator & PDF & Tune \\
\hline 
\hline
 {\sc Pythia} 8.175~\cite{Sjostrand:2007gs} & CT10~\cite{Gao:2013xoa} & AU2~\cite{ATL-PHYS-PUB-2012-003} \\
 {\sc Pythia} 8.186 & NNPDF2.3~\cite{Ball:2012cx} & Monash~\cite{Skands:2014pea} \\
 {\sc Pythia} 8.186 & NNPDF2.3 & A14~\cite{ATL-PHYS-PUB-2014-021}  \\
 {\sc Herwig++} 2.6.3~\cite{Bahr:2008pv,Arnold:2012fq} & CTEQ6L1~\cite{Stump:2003yu} & UE-EE3~\cite{Gieseke:2012ft} \\
{\sc Herwig++} 2.7.1~\cite{Bellm:2013hwb} & CTEQ6L1 & UE-EE5~\cite{Seymour:2013qka} \\
 {\sc Pythia} 6.428~\cite{Sjostrand:2006za} & CTEQ6L1 & P2012~\cite{Skands:2010ak} \\
 {\sc Pythia} 6.428 & CTEQ6L1 & P2012RadLo~\cite{Skands:2010ak} \\
 {\sc Pythia} 6.428 & CTEQ6L1 & P2012RadHi~\cite{Skands:2010ak} \\
\hline
\hline
\end{tabular}
\caption{Monte Carlo samples used in this analysis. The abbreviations ME, PDF, and UE respectively stand for matrix element, parton distribution function, and underlying event. `Tune' refers to the set of tunable MC parameters used.}
\label{tab:mc_samples1}
\end{table}

\section{Unfolding}
\label{sec:particlelevel}

\begin{figure}[h!]
\begin{center}
\includegraphics[width=0.8\textwidth]{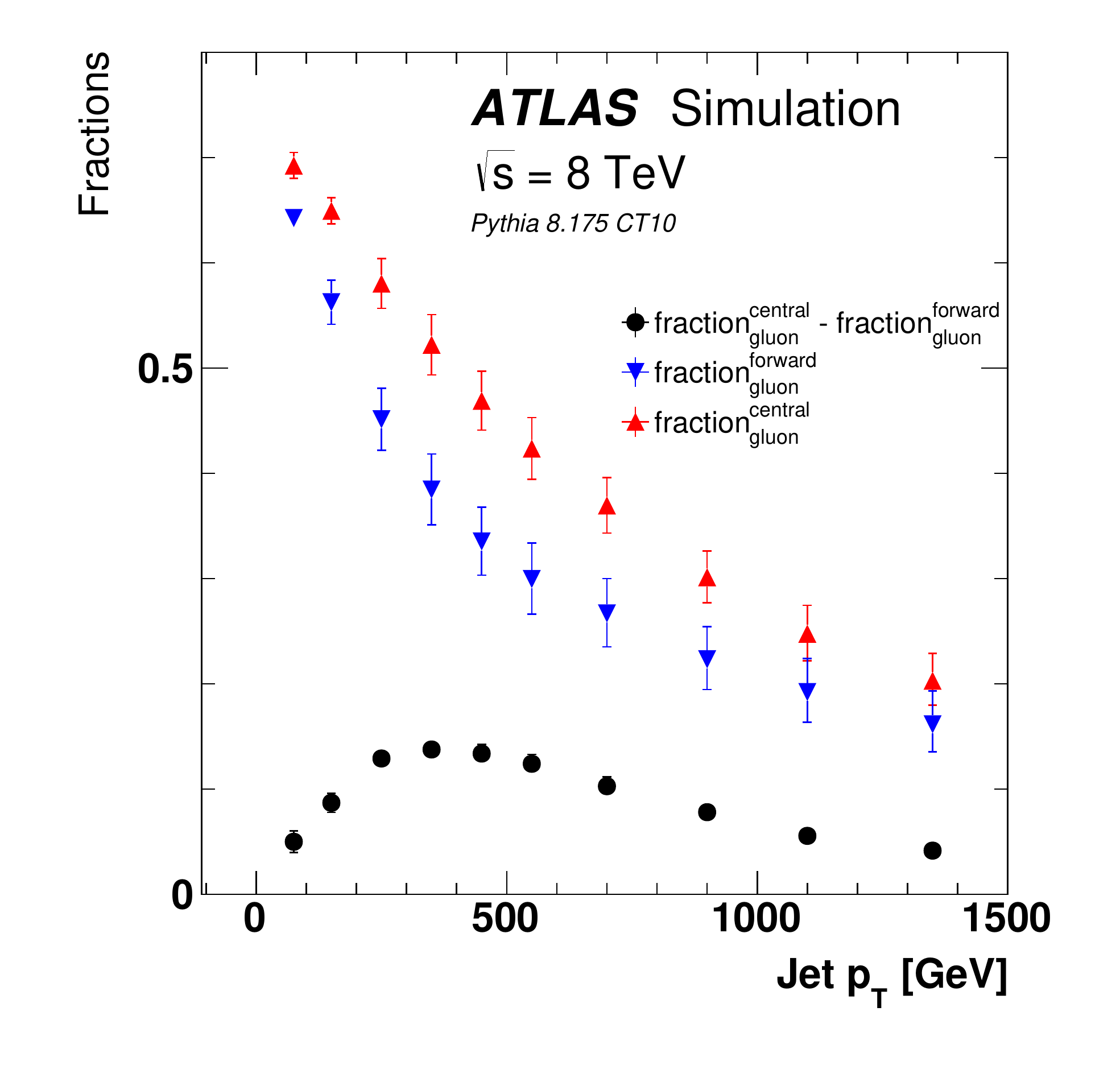}
\end{center}	
\caption{The simulated fraction of jets originating from gluons as a function of jet $p_\text{T}$ for the more forward jet (down triangle), the more central jet (up triangle), and the difference between these two fractions (circle).  The fractions are derived from {\sc Pythia} 8 with the CT10 PDF set and the error bars represent the PDF and matrix element uncertainties, further discussed in Sect.~\ref{sec:results}.  The uncertainties on the fraction difference are computed from propagating the uncertainties on the more forward and more central fractions, treating as fully correlated.}
\label{fig:qgfrac}
\end{figure}

The measurement is carried out within a fiducial volume matching the experimental selection to avoid extrapolation into unmeasured kinematic regions that have additional model dependence and related uncertainties.  Particle-level definitions of the reconstructed objects are chosen to be as close as possible to those described in Sect.~\ref{sec:atlas}.  Particle-level jets are clustered from  generated stable particles with a mean lifetime $\tau>30$~ps, excluding muons and neutrinos.  As with the detector-level jets, particle-level jets are clustered with the anti-$k_t$ $R=0.4$ algorithm.  Any charged particle clustered in a particle-level jet is considered for the charged-particle multiplicity calculation if it has $p_\text{T} > 500$ \MeV.  Events are required to have at least two jets with $|\eta|<2.1$ and $p_\text{T}>50$~\GeV~and the two highest-$p_\text{T}$ jets must satisfy the same $p_\text{T}$-balance requirement between the leading and subleading jet as at detector level ($p_\text{T}^\text{lead}/p_\text{T}^\text{sublead} < 1.5$).   The  $p_\text{T}$ symmetry requirement enriches the sample in a back-to-back topology and suppresses non-isolated jets.  In more than 70\% of events, the nearest jet in $\Delta R$ with $p_\text{T}$ > 25 GeV is the other selected jet and in less than 7\% of events, there is a jet with $p_\text{T}>25$ GeV within $\Delta R=0.8$ from one of the two selected jets.  Due to the high-energy and well-separated nature of  the selected jets, the hard-scatter quarks and gluons can be cleanly matched to the outgoing jets.  In this analysis, the type of a jet is defined as that of the highest-energy parton in simulation within a $\Delta R=0.4$ cone around the particle-jet's axis.\footnote{While it is possible to classify jets as quark- or gluon-initiated beyond leading order in $m_\text{jet}/E_\text{jet}$~\cite{Banfi:2006hf}, the classification is jet algorithm-dependent and unnecessary for the present considerations.  For the results presented in Sect.~\ref{sec:results} that rely on jet-type labelling, alternative definitions were considered and found to have a negligible impact compared to other sources of theoretical and experimental uncertainty.} Figure~\ref{fig:qgfrac} shows the fraction of gluon-initiated jets as a function of jet $p_\text{T}$ for the more forward and more central jet within the event.  The fraction of gluon-initiated jets decreases with $p_\text{T}$, but the difference between the more forward and more central jets peaks around $p_\text{T}\sim 350$ \GeV.  This difference is exploited in Sect.~\ref{sec:results} to extract separately the average quark- and gluon-initiated jet charged-particle multiplicity.

The average charged-particle multiplicity in particle-level jets is determined as a function of jet $p_\text{T}$.    An iterative Bayesian (IB) technique~\cite{D'Agostini:1994zf}
as implemented in the \texttt{RooUnfold} framework~\cite{Adye:2011gm} is used to unfold the two-dimensional charged-particle multiplicity and jet $p_\text{T}$ distribution.  In the IB unfolding technique, the number of iterations and the prior distribution are the input parameters.  The raw data are corrected using the simulation to account for events that pass the fiducial selection at detector level, but not the corresponding selection at particle level; this correction is the {\it fake factor}.  Then, the IB method iteratively applies Bayes' theorem using the
\textit{response matrix} to connect the prior distribution to the posterior distribution at each step, with the nominal {\sc {\sc Pythia}} 8.175 sample used for the initial prior distribution.
The response matrix describes the bin migrations between the particle-level and detector-level two-dimensional distribution of charged-particle multiplicity and jet $p_\text{T}$.  Although the response matrix is nearly diagonal, the resolution degrades at high $p_\text{T}$ where more bin-to-bin migrations from particle level to detector level occur.

The number of iterations in the IB method trades off unfolding bias against statistical fluctuations. An optimal value of four iterations is obtained by minimizing the bias when unfolding pseudo-data derived from {\sc {\sc Herwig++}} using a prior distribution and a response matrix derived from {\sc {\sc Pythia}} as a test of the methodology.  Lastly, unfolding applies another correction from simulation to the unfolded data to account for events passing the particle-level selection but not the detector-level selection; this correction is the {\it inefficiency factor}.  

\begin{figure}[h!]
\begin{center}
\begin{overpic}[width=0.99\textwidth]{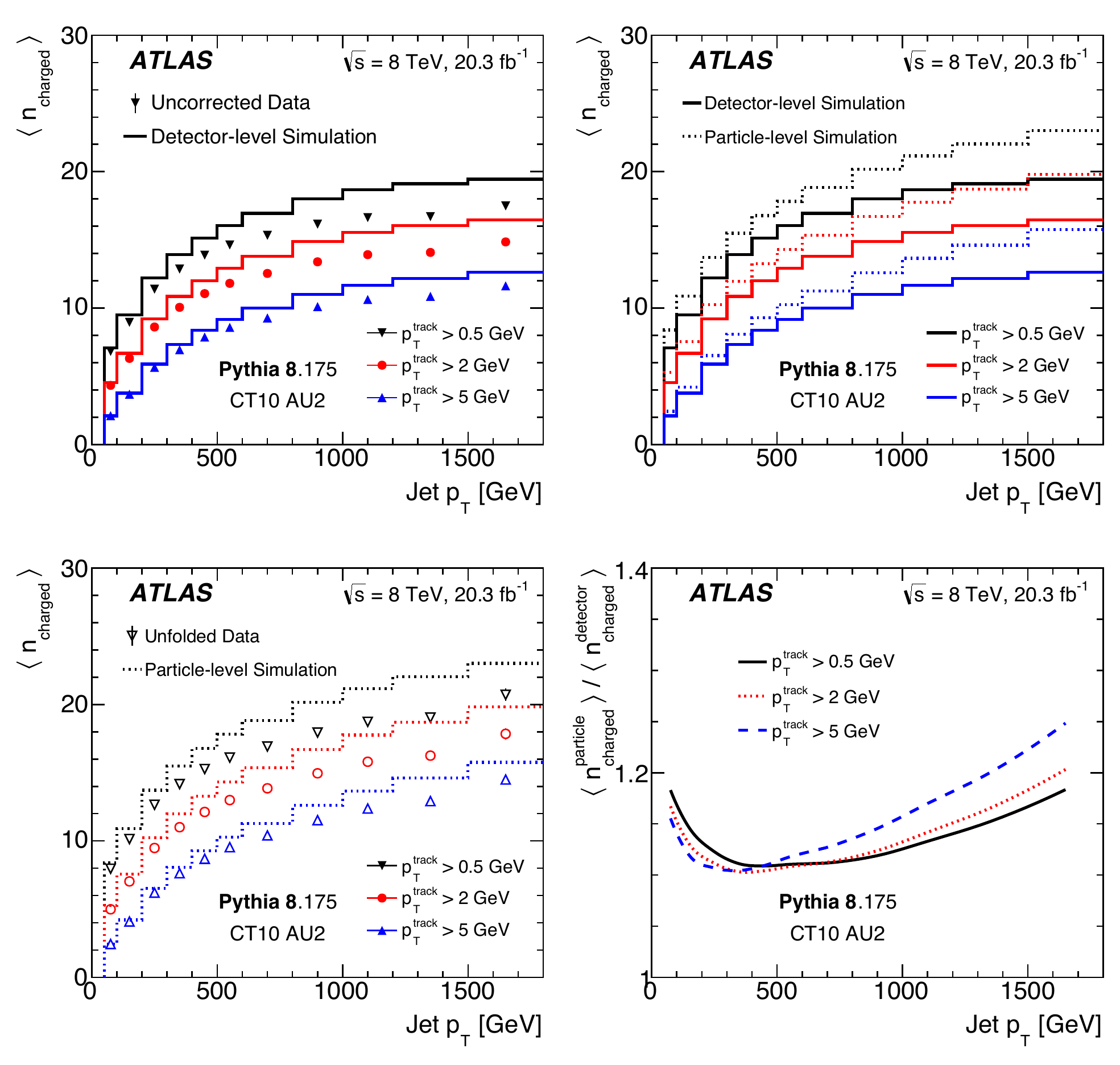}
\put(28,48){(a)}
\put(75,48){(b)}
\put(28,-2){(c)}
\put(75,-2){(d)}
\end{overpic}
\end{center}	
\caption{The jet $p_\text{T}$ dependence of (a) the average reconstructed track multiplicity for uncorrected data and detector-level simulation, (b) the average reconstructed track multiplicity for the detector-level simulation and the average charged-particle multiplicity for the particle-level simulation, (c) the average charged-particle multiplicity for the unfolded data and the particle-level simulation, and  (d) the average charged-particle multiplicity divided by the average reconstructed track multiplicity in simulation.  Three charged-particle and track $p_\text{T}$ thresholds are used in each case: 0.5 \GeV, 2 \GeV, and 5 \GeV.  {\sc Pythia 8} with the CT10 PDF and the AU2 tune are used for the simulation.  For the data, only statistical uncertainties are included in the error bars (which are smaller than the markers for most bins).}
\label{fig:truthreco}
\end{figure}
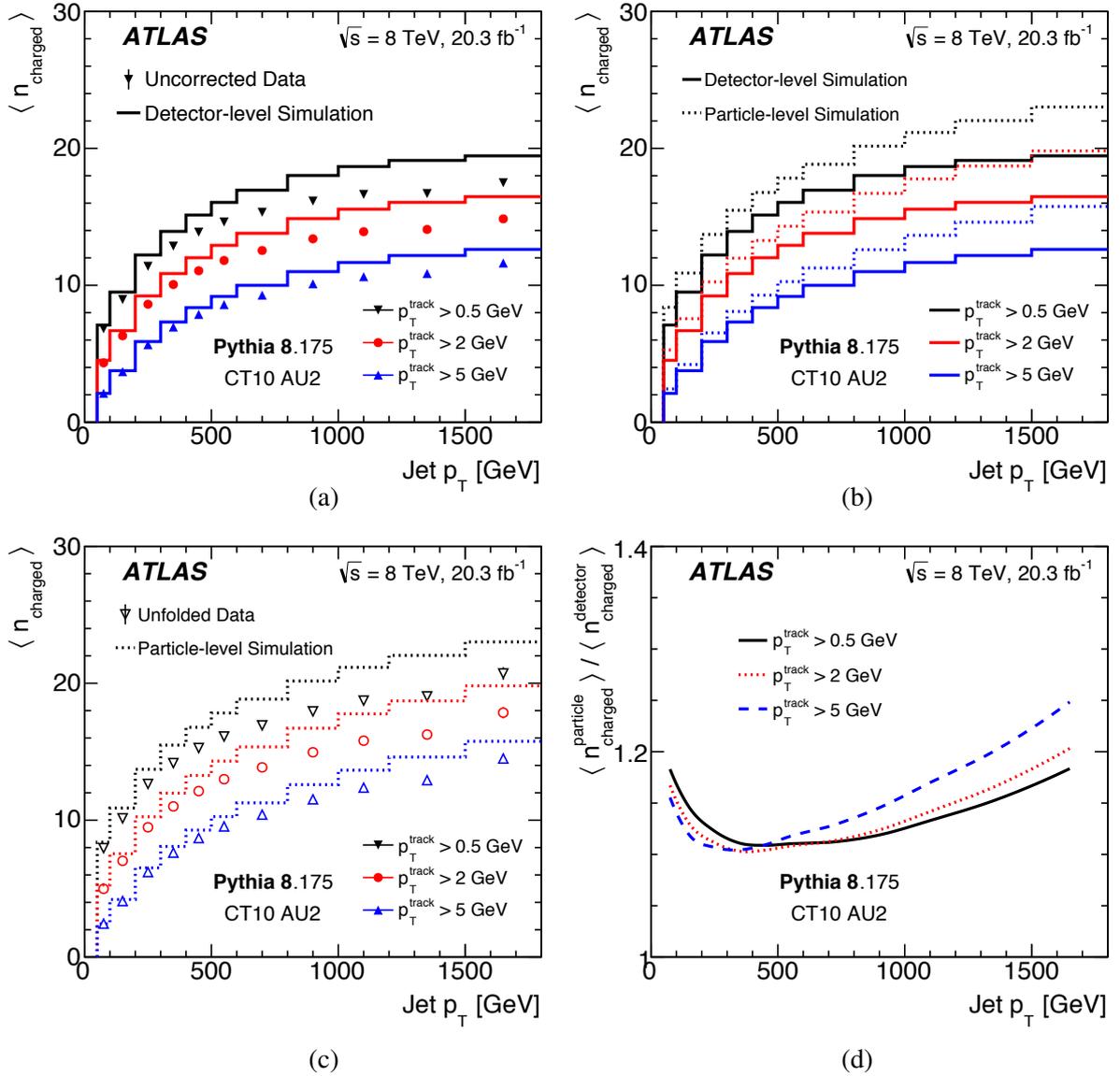

Figure~\ref{fig:truthreco} displays the $p_\text{T}$ dependence of the average charged-particle multiplicity for uncorrected data and detector-level simulation and for particle-level simulation as well as the unfolded data.  The prediction from {\sc Pythia} 8 with the AU2 tune has too many tracks compared with the uncorrected data, and the size of the data/MC difference increases with decreasing track $p_\text{T}$ threshold (Fig.~\ref{fig:truthreco}(a)).  The difference between the detector-level and particle-level simulation in Fig.~\ref{fig:truthreco}(b) (for which the ratio is given in Fig.~\ref{fig:truthreco}(d)) gives an indication of the corrections required to account for detector acceptance and resolution effects in the unfolding procedure.  Particle-level distributions in Fig.~\ref{fig:truthreco}(c) show similar trends to the detector-level ones in Fig.~\ref{fig:truthreco}(a).

\section{Systematic uncertainties}
\label{sec:uncerts}

All stages of the charged-particle multiplicity measurement are sensitive to sources of potential bias.  The three stages of the measurement are listed below, with an overview of the systematic uncertainties that impact the results at each stage:

\begin{description}
\item[Response Matrix:] For events that pass both the detector-level and particle-level fiducial selections, the response matrix describes migrations between bins when moving between the detector level and the particle level.  The response matrix is taken from simulation and various experimental uncertainties in the charged-particle multiplicity and jet $p_\text{T}$ spectra result in uncertainties in the matrix.  These uncertainties can be divided into two classes: those impacting the calorimeter-based jet $p_\text{T}$ and those impacting track reconstruction inside jets.   The dominant uncertainty at high jet $p_\text{T}$ is due to the loss of charged-particle tracks in the jet core due to track merging.  This charged energy loss uncertainty is estimated using the data/MC differences in the ratio of the track-based jet $p_\text{T}$ to the calorimeter-based jet $p_\text{T}$~\cite{Aad:2015cua}.  More charged energy is lost in the data than in the MC and thus this uncertainty is one-sided.  There are other tracking uncertainties in the track momentum scale and resolution, the track reconstruction efficiency, and the rate of tracks formed from random combinations of hits (fake tracks).  The prescription for these sub-dominant tracking uncertainties is identical to Ref.~\cite{Aad:2015cua}.  The uncertainties related to the calorimeter-based jet are sub-dominant (except in the lowest $p_\text{T}$ bins) and are due to the uncertainty in the jet energy scale and the jet energy resolution.
\item[Correction Factors:] Fake and inefficiency factors are derived from simulation to account for the fraction of events that pass either the detector-level or particle-level fiducial selection ($p_\text{T}>50$ \GeV\, $|\eta|<2.1$, and $p_\text{T}^\text{lead}/p_\text{T}^\text{sublead} < 1.5$), but not both.  These factors are derived in bins of jet $p_\text{T}$ and charged particle multiplicity, separately for the more forward and more central jets.  They are generally between $0.9$ and $1.0$ except in the first jet-$p_\text{T}$ interval (50~$<p_\text{T}<100$~\GeV), where threshold effects cause the correction factors to take values down to 0.8.  Experimental uncertainties correlated with the detector-level selection acceptance, such as the jet energy scale uncertainty, result in uncertainties in these correction factors.  Another source of uncertainty in the correction factors is the explicit dependence on the particle-level multiplicity and jet $p_\text{T}$ spectrum.  A comparison of particle-level models ({\sc {\sc Pythia}} and {\sc {\sc Herwig++}}) is used to estimate the impact on the correction factors.
\item[Unfolding Procedure:] A data-driven technique is used to estimate the potential bias from a given choice of a prior distribution and number of iterations in the IB method~\cite{Malaescu:2009dm}.  The particle-level spectrum is reweighted so that the simulated detector-level spectrum, from propagating the reweighted particle-level spectrum through the response matrix, has significantly improved agreement with the uncorrected data.  The modified detector-level distribution is unfolded with the nominal response matrix and the difference between this and the reweighted particle-level spectrum is an indication of the bias due to the unfolding method (in particular, the choice of a prior distribution).
\end{description}

A summary of the systematic uncertainties can be found in Table~\ref{tab:systs_all} and more detail about the evaluation of each uncertainty can be found in Ref.~\cite{Aad:2015cua}.  The response matrix uncertainty shown in Table~\ref{tab:systs_all} is decomposed into four categories, as described above.

\setlength\tabcolsep{1.1mm}

\begin{table}[h]
\begin{tabular}{lclcllllllll}
 {\bf Average $n_\text{charged}$} & \multicolumn{10}{c}{Jet $p_\text{T}$ Range {[}100 GeV{]}} \\
\begin{tabular}[c]{@{}c@{}}Systematic\\[-3mm]Uncertainty [\%]\end{tabular} & {[}0.5,1{]}          & {[}1,2{]}             & {[}2,3{]}            & {[}3,4{]}             & {[}4,5{]}             & {[}5,6{]}             & {[}6,8{]}             & {[}8,10{]}            & {[}10,12{]}           & {[}12,15{]}       & {[}15,18{]}     \\ \hline
Response Matrix\\\hspace{5mm}Total Jet Energy Scale & \multicolumn{1}{c}{${}^{+  1.9}_{-  1.9}$}
& \multicolumn{1}{c}{${}^{+  0.7}_{-  0.9}$}
& \multicolumn{1}{c}{${}^{+  0.6}_{-  0.8}$}
& \multicolumn{1}{c}{${}^{+  0.8}_{-  0.7}$}
& \multicolumn{1}{c}{${}^{+  0.7}_{-  0.7}$}
& \multicolumn{1}{c}{${}^{+  0.6}_{-  0.7}$}
& \multicolumn{1}{c}{${}^{+  0.6}_{-  0.7}$}
& \multicolumn{1}{c}{${}^{+  0.6}_{-  0.5}$}
& \multicolumn{1}{c}{${}^{+  0.4}_{-  0.4}$}
& \multicolumn{1}{c}{${}^{+  0.3}_{-  0.3}$}
& \multicolumn{1}{c}{${}^{+  0.8}_{-  0.7}$}
\\
\hspace{5mm}Jet Energy Resolution & \multicolumn{1}{c}{${}^{+  0.6}_{-  0.6}$}
& \multicolumn{1}{c}{${}^{+  0.1}_{-  0.1}$}
& \multicolumn{1}{c}{${}^{+  0.1}_{-  0.1}$}
& \multicolumn{1}{c}{${}^{+  0.1}_{-  0.1}$}
& \multicolumn{1}{c}{${}^{+  0.1}_{-  0.1}$}
& \multicolumn{1}{c}{${}^{+  0.1}_{-  0.1}$}
& \multicolumn{1}{c}{${}^{+  0.1}_{-  0.1}$}
& \multicolumn{1}{c}{${}^{+  0.2}_{-  0.2}$}
& \multicolumn{1}{c}{${}^{+  0.1}_{-  0.1}$}
& \multicolumn{1}{c}{${}^{+  0.1}_{-  0.1}$}
& \multicolumn{1}{c}{${}^{+  0.2}_{-  0.2}$}
\\
\hspace{5mm}Charged Energy Loss & \multicolumn{1}{c}{${}^{+  0.0}_{-  0.0}$}
& \multicolumn{1}{c}{${}^{+  0.0}_{-  0.0}$}
& \multicolumn{1}{c}{${}^{+  0.0}_{-  0.0}$}
& \multicolumn{1}{c}{${}^{+  0.0}_{-  0.0}$}
& \multicolumn{1}{c}{${}^{+  1.2}_{-  0.0}$}
& \multicolumn{1}{c}{${}^{+  1.1}_{-  0.0}$}
& \multicolumn{1}{c}{${}^{+  1.1}_{-  0.0}$}
& \multicolumn{1}{c}{${}^{+  1.1}_{-  0.0}$}
& \multicolumn{1}{c}{${}^{+  1.0}_{-  0.0}$}
& \multicolumn{1}{c}{${}^{+  3.6}_{-  0.0}$}
& \multicolumn{1}{c}{${}^{+  3.3}_{-  0.0}$}
\\
\hspace{5mm}Other Tracking & \multicolumn{1}{c}{${}^{+  1.2}_{-  0.0}$}
& \multicolumn{1}{c}{${}^{+  1.0}_{-  0.0}$}
& \multicolumn{1}{c}{${}^{+  0.9}_{-  0.0}$}
& \multicolumn{1}{c}{${}^{+  0.8}_{-  0.0}$}
& \multicolumn{1}{c}{${}^{+  0.8}_{-  0.0}$}
& \multicolumn{1}{c}{${}^{+  0.7}_{-  0.0}$}
& \multicolumn{1}{c}{${}^{+  0.7}_{-  0.0}$}
& \multicolumn{1}{c}{${}^{+  0.7}_{-  0.0}$}
& \multicolumn{1}{c}{${}^{+  0.7}_{-  0.0}$}
& \multicolumn{1}{c}{${}^{+  0.7}_{-  0.0}$}
& \multicolumn{1}{c}{${}^{+  0.8}_{-  0.0}$}
\\
Correction Factors & \multicolumn{1}{c}{${}^{+  0.1}_{-  0.1}$}
& \multicolumn{1}{c}{${}^{+  0.1}_{-  0.1}$}
& \multicolumn{1}{c}{${}^{+  0.1}_{-  0.1}$}
& \multicolumn{1}{c}{${}^{+  0.1}_{-  0.1}$}
& \multicolumn{1}{c}{${}^{+  0.1}_{-  0.1}$}
& \multicolumn{1}{c}{${}^{+  0.1}_{-  0.1}$}
& \multicolumn{1}{c}{${}^{+  0.0}_{-  0.0}$}
& \multicolumn{1}{c}{${}^{+  0.0}_{-  0.0}$}
& \multicolumn{1}{c}{${}^{+  0.0}_{-  0.0}$}
& \multicolumn{1}{c}{${}^{+  0.0}_{-  0.0}$}
& \multicolumn{1}{c}{${}^{+  0.0}_{-  0.0}$}
\\
Unfolding Procedure& \multicolumn{1}{c}{${}^{+  6.4}_{-  6.4}$}
& \multicolumn{1}{c}{${}^{+  3.4}_{-  3.4}$}
& \multicolumn{1}{c}{${}^{+  0.6}_{-  0.6}$}
& \multicolumn{1}{c}{${}^{+  0.8}_{-  0.8}$}
& \multicolumn{1}{c}{${}^{+  0.6}_{-  0.6}$}
& \multicolumn{1}{c}{${}^{+  0.4}_{-  0.4}$}
& \multicolumn{1}{c}{${}^{+  0.4}_{-  0.4}$}
& \multicolumn{1}{c}{${}^{+  0.2}_{-  0.2}$}
& \multicolumn{1}{c}{${}^{+  0.2}_{-  0.2}$}
& \multicolumn{1}{c}{${}^{+  0.2}_{-  0.2}$}
& \multicolumn{1}{c}{${}^{+  0.1}_{-  0.1}$}
\\
\hline
Total Systematic & \multicolumn{1}{c}{${}^{+  6.8}_{-  6.7}$}
& \multicolumn{1}{c}{${}^{+  3.6}_{-  3.5}$}
& \multicolumn{1}{c}{${}^{+  1.2}_{-  1.0}$}
& \multicolumn{1}{c}{${}^{+  1.4}_{-  1.1}$}
& \multicolumn{1}{c}{${}^{+  1.7}_{-  0.9}$}
& \multicolumn{1}{c}{${}^{+  1.5}_{-  0.8}$}
& \multicolumn{1}{c}{${}^{+  1.5}_{-  0.8}$}
& \multicolumn{1}{c}{${}^{+  1.4}_{-  0.6}$}
& \multicolumn{1}{c}{${}^{+  1.3}_{-  0.5}$}
& \multicolumn{1}{c}{${}^{+  3.7}_{-  0.4}$}
& \multicolumn{1}{c}{${}^{+  3.5}_{-  0.7}$}
\\
Data Statistics & \multicolumn{1}{c}{  0.5}
& \multicolumn{1}{c}{  0.2}
& \multicolumn{1}{c}{  0.1}
& \multicolumn{1}{c}{  0.1}
& \multicolumn{1}{c}{  0.0}
& \multicolumn{1}{c}{  0.1}
& \multicolumn{1}{c}{  0.1}
& \multicolumn{1}{c}{  0.3}
& \multicolumn{1}{c}{  0.6}
& \multicolumn{1}{c}{  1.2}
& \multicolumn{1}{c}{  3.9}
\\
Total Uncertainty & \multicolumn{1}{c}{${}^{+  6.8}_{-  6.7}$}
& \multicolumn{1}{c}{${}^{+  3.6}_{-  3.6}$}
& \multicolumn{1}{c}{${}^{+  1.2}_{-  1.0}$}
& \multicolumn{1}{c}{${}^{+  1.4}_{-  1.1}$}
& \multicolumn{1}{c}{${}^{+  1.7}_{-  0.9}$}
& \multicolumn{1}{c}{${}^{+  1.5}_{-  0.8}$}
& \multicolumn{1}{c}{${}^{+  1.5}_{-  0.8}$}
& \multicolumn{1}{c}{${}^{+  1.4}_{-  0.6}$}
& \multicolumn{1}{c}{${}^{+  1.5}_{-  0.8}$}
& \multicolumn{1}{c}{${}^{+  3.9}_{-  1.3}$}
& \multicolumn{1}{c}{${}^{+  5.2}_{-  4.0}$}
\\
\hline\hline
Measured Value& \multicolumn{1}{c}{ 7.87}
& \multicolumn{1}{c}{ 9.87}
& \multicolumn{1}{c}{12.19}
& \multicolumn{1}{c}{13.54}
& \multicolumn{1}{c}{14.59}
& \multicolumn{1}{c}{15.41}
& \multicolumn{1}{c}{16.28}
& \multicolumn{1}{c}{17.41}
& \multicolumn{1}{c}{18.25}
& \multicolumn{1}{c}{18.71}
& \multicolumn{1}{c}{20.78}
\\
\end{tabular}
\caption{A summary of all the systematic uncertainties and their impact on the $n_\text{track}$ mean for $p_\text{T}^\text{track}>0.5$ GeV and the more forward jet.  Uncertainties are given in percent.  The last row is the measured average charged particle multiplicity.  A value of 0.0 is quoted if the uncertainty is below 0.05\%.}
\label{tab:systs_all}
\end{table}

\clearpage
\newpage

\section{Results}
\label{sec:results}

\begin{figure}[h!]
\begin{center}
\subfloat[]{\includegraphics[width=0.5\textwidth]{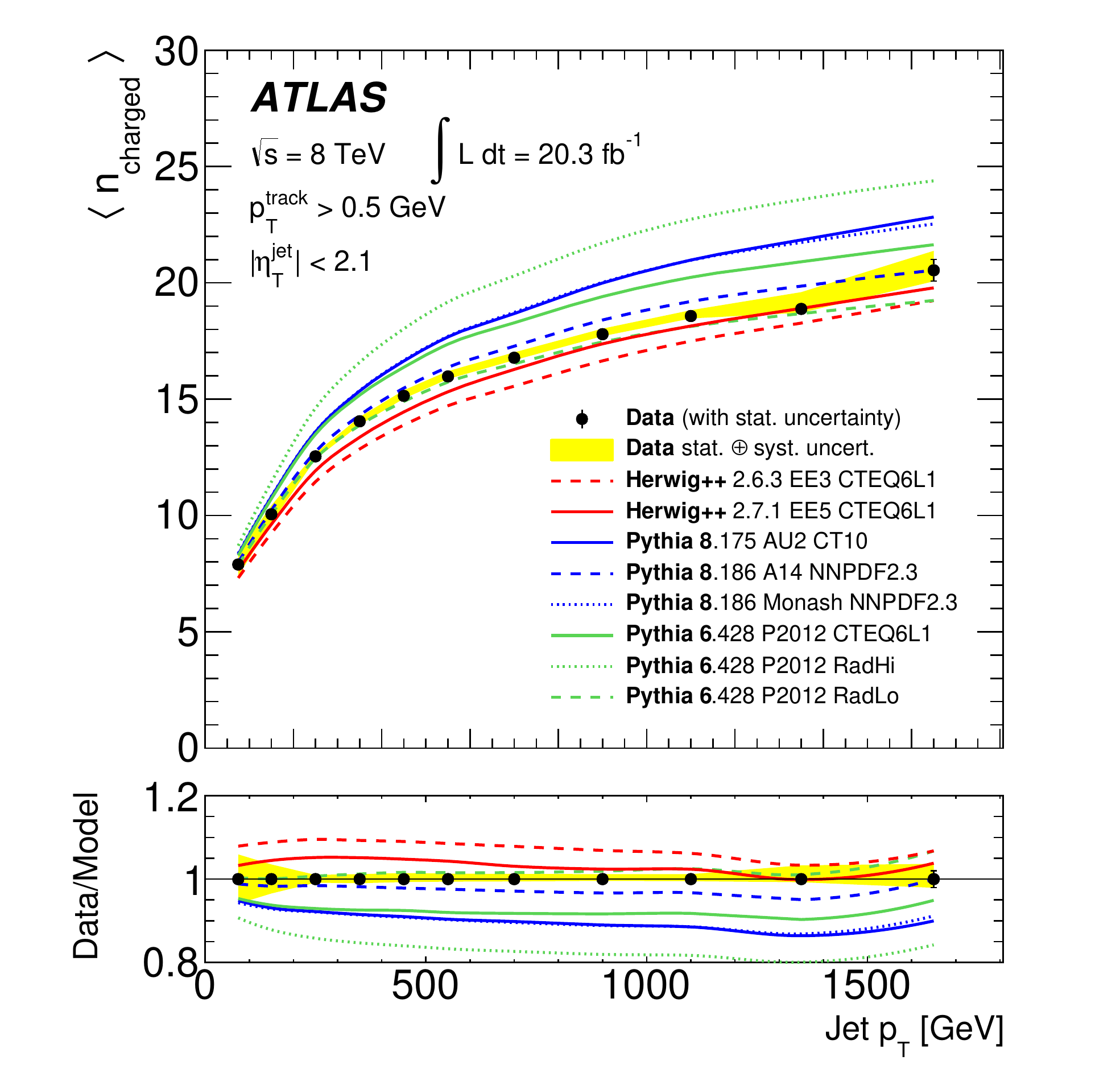}
}
\subfloat[]{
\includegraphics[width=0.5\textwidth]{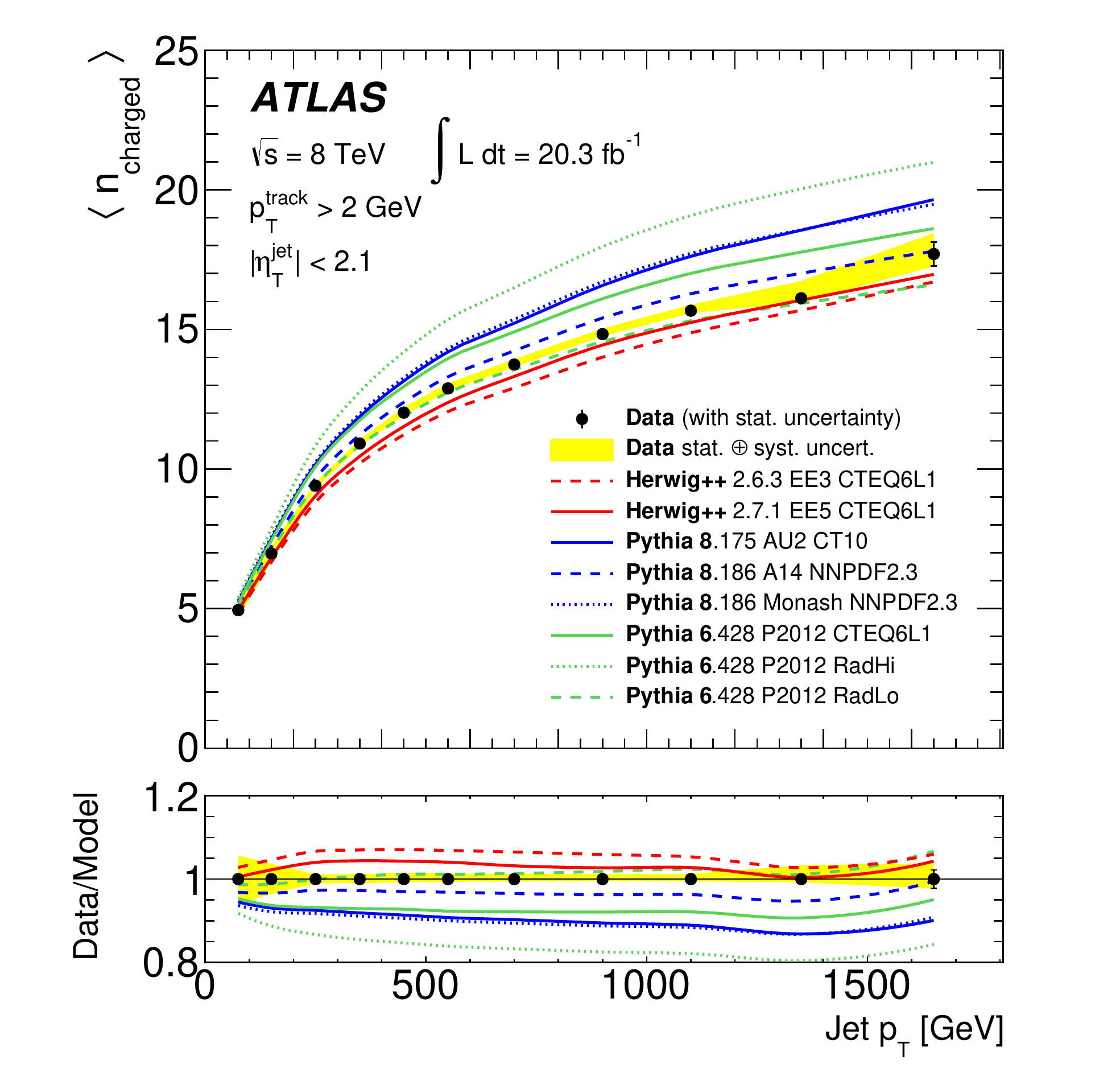}
}\\
\subfloat[]{
\includegraphics[width=0.5\textwidth]{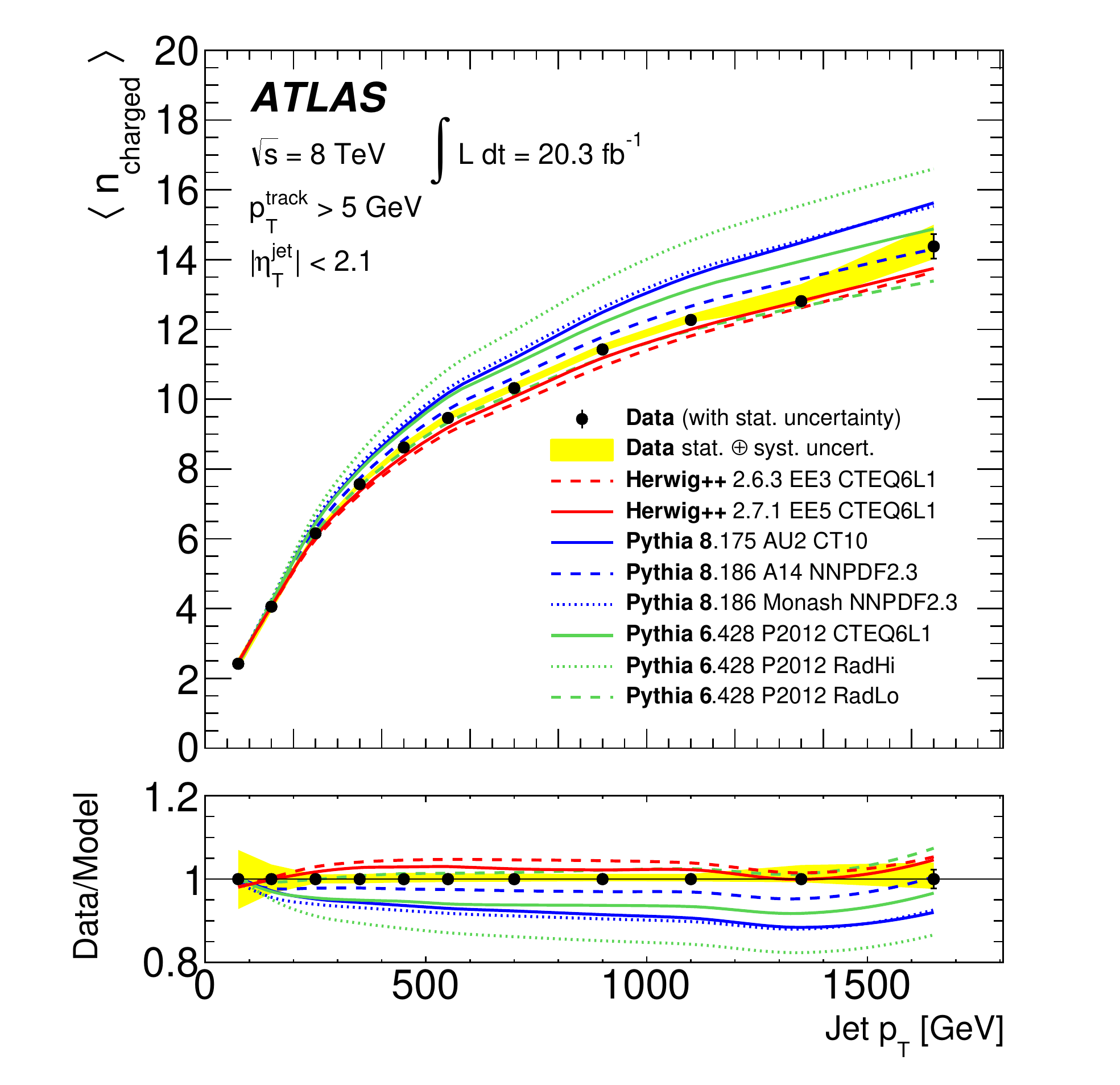}
}
\end{center}	
\caption{The measured average charged-particle multiplicity as a function of the jet $p_\text{T}$, combining the more forward and the more central jets for (a) $p_\text{T}^\text{track}>0.5$ \GeV,  (b) $p_\text{T}^\text{track}>2$ \GeV, and (c) $p_\text{T}^\text{track}>5$ \GeV.  The band around the data is the sum in quadrature of the statistical and systematic uncertainties.  Error bars on the data points represent the statistical uncertainty (which are smaller than the markers for most bins). }
\label{fig:result1}
\end{figure}

The unfolded average charged-particle multiplicity combining both the more forward and the more central jets is shown in Fig.~\ref{fig:result1}, compared with various model predictions.  As was already observed for the reconstructed data in Fig.~\ref{fig:tracks}, the average charged-particle multiplicity in data falls between the predictions of {\sc Pythia 8} and {\sc Herwig++}, independently of the underlying-event tunes.  The {\sc Pythia 8} predictions are generally higher than the data and this is more pronounced at higher jet $p_\text{T}$.  The default ATLAS tune in Run 1 (AU2) performs similarly to the Monash tune, but the prediction with A14 (the ATLAS default for the analysis of Run 2 data) is significantly closer to the data.  A previous ATLAS measurement~\cite{Aad:2011gn} of charged-particle multiplicity inside jets was included in the tuning of A14, but the jets in that measurement have $p_\text{T}\lesssim 50$ \GeV.  One important difference between A14 and Monash is that the value of $\alpha_\text{s}$ governing the amount of final-state radiation is about 10\% lower in A14 than in Monash.  This parameter has a large impact on the average charged-particle multiplicity, which is shown by the {\sc Pythia 6} lines in Fig.~\ref{fig:result1} where the Perugia radHi and radLo tunes are significantly separated from the central P2012 tune.  The $\alpha_\text{s}$ value that regulates final-state radiation is changed by factors of one half and two for these tunes with respect to the nominal Perugia 2012 tune.  The recent (and Run 2 default) EE5 underlying-event tune for {\sc Herwig++} improves the modelling of the average charged-particle multiplicity with respect to the EE3 tune (Run 1 default).

\clearpage

\begin{figure}[h!]
\begin{center}
\subfloat[]{
\includegraphics[width=0.5\textwidth]{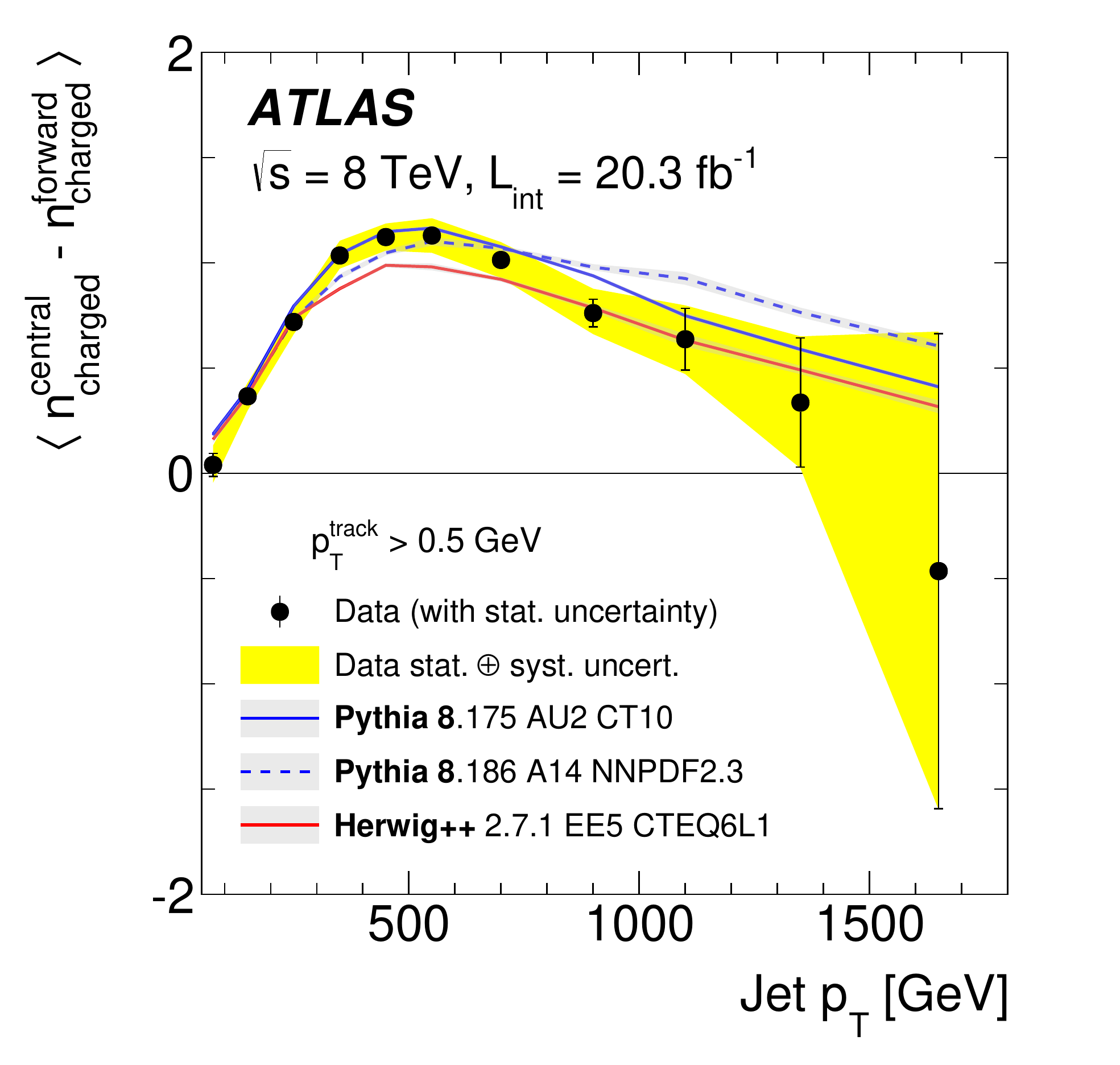}
}
\subfloat[]{
\includegraphics[width=0.5\textwidth]{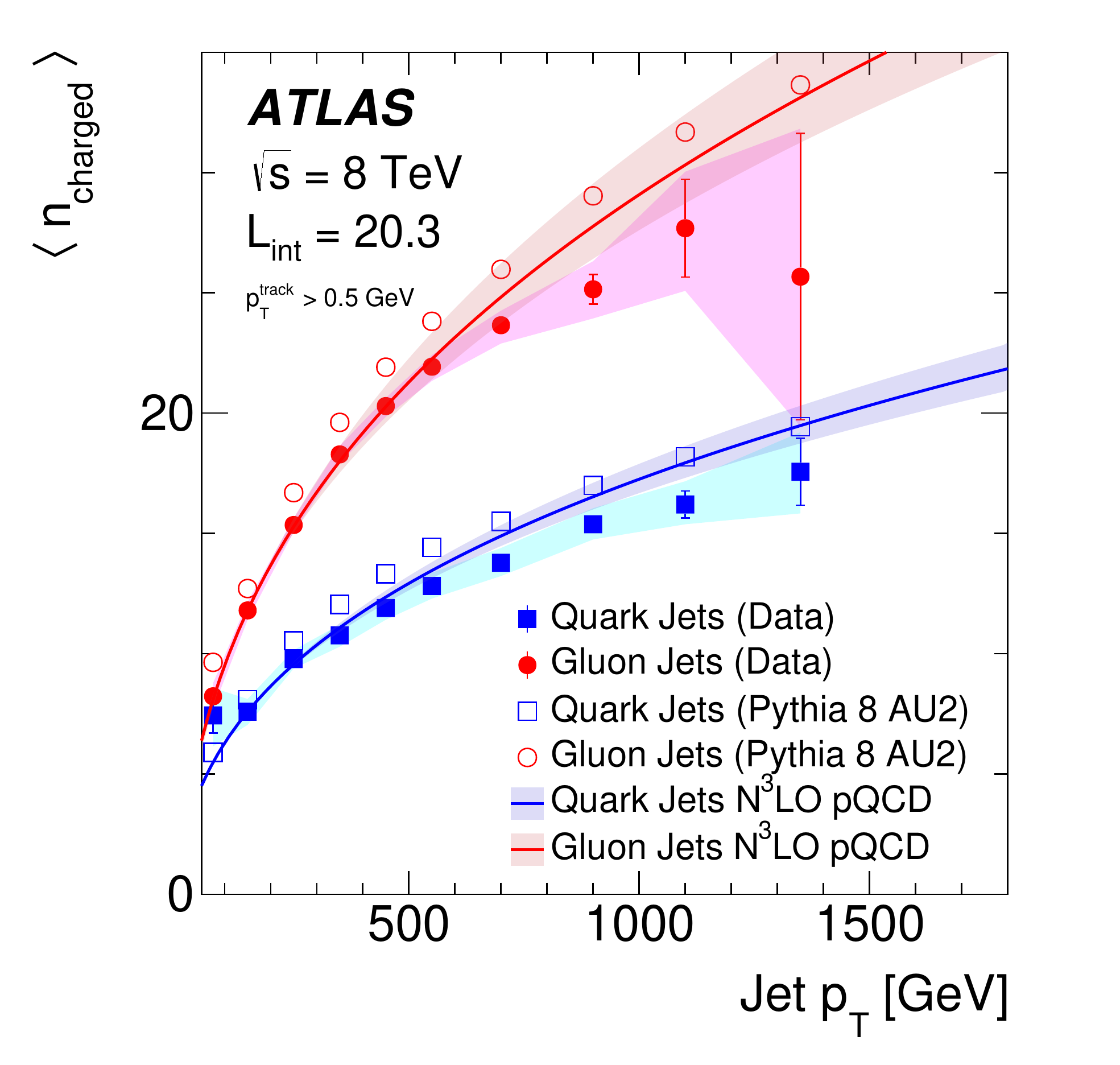}
}
\end{center}	
\caption{The jet $p_\text{T}$ dependence of (a) the difference in the average charged-particle multiplicity ($p_\text{T}^\text{track}>0.5$ \GeV) between the more forward and the more central jet.  The band for the data is the sum in quadrature of the systematic and statistical uncertainties and the error bars on the data points represent the statistical uncertainty.  Bands on the simulation include MC statistical uncertainty.  The jet $p_\text{T}$ dependence of (b) the average charged-particle multiplicity ($p_\text{T}^\text{track}>0.5$ \GeV) for quark- and gluon-initiated jets, extracted with the gluon fractions from {\sc Pythia} 8.175 with the CT10 PDF.  In addition to the experimental uncertainties, the error bands include uncertainties in the gluon fractions from both the PDF and ME uncertainties.  The MC statistical uncertainties on the open markers are smaller than the markers.  The uncertainty band for the N${}^3$LO pQCD prediction is determined by varying the scale $\mu$ by a factor of two up and down.  The markers are truncated at the penultimate $p_\text{T}$ bin in the right because within statistical uncertainty, the more forward and more central jet constituent charged-particle multiplicities are consistent with each other in the last bin.}
\label{fig:result2}
\end{figure}

The difference in the average charged-particle multiplicity between the more forward and the more central jet is sensitive to the difference between quark and gluon constituent multiplicities.   Figure~\ref{fig:result2}(a) shows that the difference is significant for $p_\text{T}\lesssim 1.1$ \TeV.  The shape is governed by the difference in the gluon fraction between the more forward and the more central jet, which was shown in Fig.~\ref{fig:qgfrac} to peak around $p_\text{T}\sim 350$ \GeV.  The average difference, combined with the gluon fraction, can be used to extract the average charged-particle multiplicity for quark- and gluon-initiated jets separately.   Given the quark and gluon fractions $f_{q,g}^{f,c}$ with $f=\text{more forward}$, $c=\text{more central}$, $q=\text{quark}$, $g=\text{gluon}$ and $f_{q}+f_{g}=1$, the average charged-particle multiplicity for quark- and gluon-initiated jets is extracted by solving the system of equations in Eq.~(\ref{eq:system});

\begin{align}
\label{eq:system}
\langle n_\text{charged}^f\rangle&=f_q^f\langle n_\text{charged}^q\rangle+f_g^f\langle n_\text{charged}^g\rangle\\\nonumber
\langle n_\text{charged}^c\rangle&=f_q^c\langle n_\text{charged}^q\rangle+f_g^c\langle n_\text{charged}^g\rangle.
\end{align}

Given the jet $p_\text{T}$, the charged particle multiplicity inside jets does not vary significantly with $\eta$.  This is confirmed by checking that the solution to Eq.~\ref{eq:system} reproduces the quark and gluon jet charged particle multiplicities for both {\sc Pythia 8} and {\sc Herwig++} to better than $1\%$ across most of the $p_\text{T}$ range.  The extracted $p_\text{T}$ dependence of the average charged-particle multiplicities for quark- and gluon-initiated jets is shown in Fig.~\ref{fig:result2}(b).  {\sc Pythia 8} with the CT10 PDF set is used to determine the gluon fractions.  The experimental uncertainties are propagated through Eq.~(\ref{eq:system}) by recomputing the quark and gluon average charged-particle multiplicities for each variation accounting for a systematic uncertainty; the more forward and more central jet uncertainties are treated as being fully correlated.  In addition to the experimental uncertainties, the error bands in Fig.~\ref{fig:result2}(b) include uncertainties in the gluon fractions from both the PDF and matrix element (ME) uncertainties.  The PDF uncertainty is determined using the CT10 eigenvector PDF sets and validated by comparing CT10 and NNPDF.  The ME uncertainty is estimated by comparing the fractions $f_{q,g}^{f,c}$ from {\sc Pythia 8} and {\sc Herwig++} after reweighting the {\sc Pythia 8} sample with CT10 to CTEQ6L1 to match the PDF used for {\sc Herwig++}.  All PDF re-weighting is performed using LHAPDF6~\cite{Buckley:2014ana}.  The PDF and ME uncertainties are comparable in size to the total experimental uncertainty.  As expected, the average multiplicity increases with jet $p_\text{T}$ for both the quark-initiated jets and gluon-initiated jets, but increases faster for gluon-initiated jets.  Furthermore, the multiplicity is significantly higher for gluon-initiated jets than for quark-initiated jets.  The average charged-particle multiplicity in {\sc Pythia 8} with the AU2 tune is higher than in the data for both the quark- and gluon-initiated jets.  In addition to predictions from leading-logarithm parton shower simulations, calculations of the scale dependence for the parton multiplicity inside jets have been performed in perturbative quantum chromodynamics (pQCD).  Up to a non-perturbative factor that is constant for the jet $p_\text{T}$ range considered in this analysis,\footnote{This factor is found to be about 0.19 for gluon jets and 0.25 for quark-initiated jets.} these calculations can be interpreted as a prediction for the scale dependence of $\langle n_\text{charged}\rangle$ for quark- and gluon-initiated jets.  There are further caveats to the predictability of such a calculation since $n_\text{charged}$ is not infrared safe or even Sudakov safe~\cite{Larkoski:2015lea}.  Therefore, the formal accuracy of the series expansion in $\sqrt{\alpha_\text{s}}$ is unknown.  Given these caveats, the next-to-next-to-next-to-leading-order (N${}^3$LO) pQCD calculation~\cite{Capella:1999ms,Dremin:1999ji} is overlaid in Fig.~\ref{fig:result2} with renormalization scale $\mu=Rp_\text{T}$ in the five-flavour scheme and $R=0.4$.  The theoretical error band is calculated by varying $\mu$ by a factor of two.  The prediction cannot give the absolute scale, and therefore the curve is normalized in the second $p_\text{T}$ bin ($100$ \GeV $<p_\text{T}<200$ \GeV) where the statistical uncertainty is small.  The predicted scale dependence for gluon-initiated jets is consistent with the data within the uncertainty bands while the curve for quark-initiated jets is higher than the data by about one standard deviation.

 \clearpage
 
\section{Summary}
\label{sec:summary}

This paper presents a measurement of the $p_\text{T}$ dependence of the average jet charged-particle multiplicity in dijet events from 20.3 fb${}^{-1}$ of $\sqrt{s}=8$ \TeV~$pp$ collision data recorded by the ATLAS detector at the LHC.  The measured charged-particle multiplicity distribution is unfolded to correct for the detector acceptance and resolution to facilitate direct comparison to particle-level models.  Comparisons are made at particle level between the measured average charged-particle multiplicity and various models of jet formation.  Significant differences are observed between the simulations using Run 1 tunes and the data, but the Run 2 tunes for both {\sc Pythia 8} and {\sc Herwig++} significantly improve the modelling of the average $n_\text{charge}$.  Furthermore, quark- and gluon-initiated jet constituent charged-particle multiplicities are extracted and compared with simulations and calculations.  As expected, the extracted gluon-initiated jet constituent charged-particle multiplicity is higher than the corresponding quantity for quark-initiated jets and a calculation of the $p_\text{T}$-dependence accurately models the trend observed in the data.  The particle-level spectra are available~\cite{hepdata} for further interpretation and can serve as a benchmark for future measurements of the evolution of non-perturbative jet observables to validate MC predictions and tune their model parameters.

%-------------------------------------------------------------------------------
\section*{Acknowledgements}
%-------------------------------------------------------------------------------

% Acknowledgements for papers with collision data
% Version 15-Sep-2015

% Standard acknowledgements start here
%----------------------------------------------
We thank CERN for the very successful operation of the LHC, as well as the
support staff from our institutions without whom ATLAS could not be
operated efficiently.

We acknowledge the support of ANPCyT, Argentina; YerPhI, Armenia; ARC, Australia; BMWFW and FWF, Austria; ANAS, Azerbaijan; SSTC, Belarus; CNPq and FAPESP, Brazil; NSERC, NRC and CFI, Canada; CERN; CONICYT, Chile; CAS, MOST and NSFC, China; COLCIENCIAS, Colombia; MSMT CR, MPO CR and VSC CR, Czech Republic; DNRF, DNSRC and Lundbeck Foundation, Denmark; IN2P3-CNRS, CEA-DSM/IRFU, France; GNSF, Georgia; BMBF, HGF, and MPG, Germany; GSRT, Greece; RGC, Hong Kong SAR, China; ISF, I-CORE and Benoziyo Center, Israel; INFN, Italy; MEXT and JSPS, Japan; CNRST, Morocco; FOM and NWO, Netherlands; RCN, Norway; MNiSW and NCN, Poland; FCT, Portugal; MNE/IFA, Romania; MES of Russia and NRC KI, Russian Federation; JINR; MESTD, Serbia; MSSR, Slovakia; ARRS and MIZ\v{S}, Slovenia; DST/NRF, South Africa; MINECO, Spain; SRC and Wallenberg Foundation, Sweden; SERI, SNSF and Cantons of Bern and Geneva, Switzerland; MOST, Taiwan; TAEK, Turkey; STFC, United Kingdom; DOE and NSF, United States of America. In addition, individual groups and members have received support from BCKDF, the Canada Council, CANARIE, CRC, Compute Canada, FQRNT, and the Ontario Innovation Trust, Canada; EPLANET, ERC, FP7, Horizon 2020 and Marie Sk?odowska-Curie Actions, European Union; Investissements d'Avenir Labex and Idex, ANR, Region Auvergne and Fondation Partager le Savoir, France; DFG and AvH Foundation, Germany; Herakleitos, Thales and Aristeia programmes co-financed by EU-ESF and the Greek NSRF; BSF, GIF and Minerva, Israel; BRF, Norway; the Royal Society and Leverhulme Trust, United Kingdom.

The crucial computing support from all WLCG partners is acknowledged
gratefully, in particular from CERN and the ATLAS Tier-1 facilities at
TRIUMF (Canada), NDGF (Denmark, Norway, Sweden), CC-IN2P3 (France),
KIT/GridKA (Germany), INFN-CNAF (Italy), NL-T1 (Netherlands), PIC (Spain),
ASGC (Taiwan), RAL (UK) and BNL (USA) and in the Tier-2 facilities
worldwide.
%----------------------------------------------

\printbibliography

\newpage
\clearpage

\newpage % ATLAS Collaboration author list
% Data extracted on 22-Dec-2015 for paper reference STDM-2015-12
%\documentclass[11pt]{article}
%\usepackage{a4wide}\begin{document}
\begin{flushleft}
{\Large The ATLAS Collaboration}

\bigskip

G.~Aad$^\textrm{\scriptsize 87}$,
B.~Abbott$^\textrm{\scriptsize 114}$,
J.~Abdallah$^\textrm{\scriptsize 152}$,
O.~Abdinov$^\textrm{\scriptsize 11}$,
B.~Abeloos$^\textrm{\scriptsize 118}$,
R.~Aben$^\textrm{\scriptsize 108}$,
M.~Abolins$^\textrm{\scriptsize 92}$,
O.S.~AbouZeid$^\textrm{\scriptsize 138}$,
N.L.~Abraham$^\textrm{\scriptsize 150}$,
H.~Abramowicz$^\textrm{\scriptsize 154}$
% ...
\bigskip
\\
% ...
$^{11}$ Institute of Physics, Azerbaijan Academy of Sciences, Baku, Azerbaijan\\
% ...
$^{87}$ CPPM, Aix-Marseille Universit{\'e} and CNRS/IN2P3, Marseille, France\\
% ...
$^{92}$ Department of Physics and Astronomy, Michigan State University, East Lansing MI, United States of America\\
% ...
$^{108}$ Nikhef National Institute for Subatomic Physics and University of Amsterdam, Amsterdam, Netherlands\\
% ...
$^{114}$ Homer L. Dodge Department of Physics and Astronomy, University of Oklahoma, Norman OK, United States of America\\
% ...
$^{118}$ LAL, Univ. Paris-Sud, CNRS/IN2P3, Universit{\'e} Paris-Saclay, Orsay, France\\
% ...
$^{138}$ Santa Cruz Institute for Particle Physics, University of California Santa Cruz, Santa Cruz CA, United States of America\\
% ...
$^{150}$ Department of Physics and Astronomy, University of Sussex, Brighton, United Kingdom\\
% ...
$^{152}$ Institute of Physics, Academia Sinica, Taipei, Taiwan\\
% ...
$^{154}$ Raymond and Beverly Sackler School of Physics and Astronomy, Tel Aviv University, Tel Aviv, Israel\\
% ...
\end{flushleft}

\end{document}